# Temporal Correlation between Ionospheric Storm-Enhanced Density Plume and Plasmaspheric Plume Occurrence

**Xiangning Chu[1], Shunrong Zhang[2], Jacob Bortnik[3], Naomi Maruyama[1], Jerry Goldstein[45], Kausik Chatterjee[167], David Malaspina[18], Phil Erickson[2]**

[1] Laboratory for Atmospheric and Space Physics, University of Colorado Boulder, Boulder, Colorado, USA

[2] Haystack Observatory, Massachusetts Institute of Technology, Westford, MA, United States

[3] Department of Atmospheric and Oceanic Sciences, University of California, Los Angeles, CA, USA

[4] Space Science Department, Southwest Research Institute, San Antonio, TX, USA

[5] Department of Physics and Astronomy, University of Texas at San Antonio, San Antonio, TX, USA

[6] Stochastic Research Institute, Chantilly, VA

[7] University of New Mexico, Albuquerque, NM

[8] Astrophysical and Planetary Sciences Department, University of Colorado, Boulder, CO, USA; David.Malaspina@colorado.edu

Corresponding author: Xiangning Chu (chuxiangning@gmail.com)

**Key Points:**

- Plasmapause asymmetry (PPA) index, developed using a machine learning approach, enables identification of plasmaspheric plumes
- Temporal alignment between start and end times of storm-enhanced density (SED) and plasmaspheric plumes is found for 75 magnetic storms
- Plume and SED formation are regulated by peaks in solar wind Ey, direct-driven electric fields, SAPS, and substorm-related electric fields

## Abstract

We present a statistical evidence of a temporal correlation between ionospheric storm-enhanced density (SED) and plasmaspheric plumes during magnetic storms. We identified SED plumes in 75 storms between 2010-2024 with sufficient total electron content (TEC) coverage. We also identified plasmaspheric plumes during these storms using a plasmapause asymmetry (PPA) index derived from the machine-learning-based DEN3D model. We found that all SED plumes coincided with plasmaspheric plumes. The start and end times of both phenomena exhibit near-zero lag within a half-hour delay, confirming their synchronized evolution. For the first time, superposed epoch analyses reveal that plume onsets align with solar wind driving and geomagnetic activity, including peaks in the electric field, coupling function, auroral electrojet indices, and the asymmetric ring current index, all of which facilitate plume formation. This work provides timely new insights into magnetosphere-ionosphere coupling processes and demonstrates the utility of machine-learning-based diagnostics for space physics/weather.

## Plain Language Summary

During magnetic storms, two key space weather phenomena occur: storm-enhanced density (SED) plumes in the ionosphere (Earth’s ionized upper atmosphere) and plasmaspheric plumes in the plasmasphere (a toroidal-shaped region of plasma that escapes the ionosphere and becomes trapped by the Earth’s magnetic field). In the magnetosphere, eroded plumes, a global column of cold plasma eroded and transported sunward, can influence spacecraft charging and enhance energetic particle populations, posing risks to space-based assets. In the ionosphere, SED plume, a dense and poleward-extending channel of plasma with sharp electron density gradients, can disrupt satellite navigation and communications. In this study, we use a machine learning model to track plasmaspheric plumes and compare their timing with SED events identified from global maps of total electron content. We find a strong one-to-one match: plasmaspheric plumes and ionospheric SED plumes occur concurrently during magnetic storms, with closely aligned start and end times. This correspondence suggests that the two phenomena are linked through electric fields and plasma flows along magnetic field lines. We also show that changes in the solar wind can trigger plumes and SEDs. These results improve understanding of how Earth’s space environment responds to solar storms and support future space weather forecasting tools.

## 1 Introduction

### 1.1 Background

Geomagnetic storms significantly restructure the near-Earth plasma environment, affecting both the magnetosphere and the ionosphere. In the magnetosphere, enhanced convection erodes the outer plasmasphere, stripping cold and dense plasma from the nightside and subsequently driving cold plasma along reconfigured equal-potentials toward the dayside magnetopause, and forming sunward-directed plasmaspheric plumes (Chen and Grebowsky, 1974; Ober et al., 1997; Sandel et al., 2001; Darrouzet et al., 2008; Garcia et al., 2003; Moldwin et al., 2004; Goldstein, 2003, 2005). In the ionosphere, storm-time electrodynamics contribute to the generation of storm-enhanced density (SED) plumes, narrow channels of elevated Global Navigation Satellite Systems (GNSS) total electron content (TEC) in the afternoon midlatitude ionosphere extend sunward and poleward, often evolving into tongues of ionization in the polar cap (e.g., Foster, 1993, 2021). Identifying when and where SEDs and plasmaspheric plumes coincide, and the conditions that drive their linkage, advances understanding of a fundamental aspect of storm-time magnetosphere--ionosphere coupling.

### 1.2 State of knowledge and gaps

With conjugate observations of global GNSS TEC maps and IMAGE EUV images during a magnetic storm, Foster et al. (2002) show that time snapshots of an ionospheric SED plume maps closely to a plasmaspheric plume, forming the broader concept of Geospace plume (Foster et al., 2020). Single-time snapshots of plasmaspheric images and ionospheric TEC maps from case studies show that SED and plasmaspheric plumes roughly colocate in the same local time sector along magnetic field lines, together with corresponding electric fields and plasma flows, suggesting a direct magnetosphere-ionosphere connection that may be enabled or facilitated by coupling and feekback (Foster et al., 2002; 2005; 2007; Goldstein et al., 2003; Moldwin et al., 2016; Zou et al., 2014; 2021; Aa et al., 2003; 2024; Li et al., 2023; Yizengaw et al., 2006; 2008). SEDs are usually considered to be related to storm-time penetration electric fields (Heelis et al., 2009; Zou et al., 2014; Liu et al., 2016; Zhang and Aa, 2021;), along with related subauroral and/or mid-latitude ionospheric electrodynamics forcing and thermospheric dynamics driven by subauroral polarization streams (SAPS) fast flows (Foster et al., 2002, 2007, 2020; Moldwin et al., 2016; Yizengaw et al., 2006), equatorward neutral winds (Anderson, 1976; Mendillo, 2006; Balan et al., 2010;), thermospheric composition effects (Immel et al., 2001; Wang et al., 2012), and poleward expansion of the equatorial ionization anomaly (Kelley et al., 2004; Mannucci et al., 2005; Tsurutani et al., 2004).

A limited statistical study of 31 events in the first half year of 2001 found that when a plasmaspheric plume was observed, an SED was present in 100% of cases during the same storm which were located in the American sector with sufficient TEC coverage. However, this correspondence was lower in other longitude regions potentially due to factors including seasonal variations, geographic and geomagnetic pole offsets (Foster and Erickson, 2013) and less densely sampled TEC coverage, namely 50% in the European sector, and 20% in the Asian sector (Yizengaw et al., 2008). In general, temporal correspondences reported in the literature are often based primarily on isolated time snapshots and case studies of storms, posing difficulties with determining the overall strength of the plasmaspheric and SED plume connection. The questions remain open: (1) whether a one-to-one correspondence consistently holds between plasmasphere and SED plumes across magnetic storms with sufficient observational coverage, and (2) whether their start and end times show systematic offsets. In particular, the temporal evolution of this linkage over storms has not been systematically quantified with a large statistical dataset.

Moreover, the solar wind driving and geomagnetic activity that initiate and sustain the dynamics of SED and plasmaspheric plumes remain poorly understood. More SED plumes were observed during the solar maximum and declining phases than during the solar minimum (Aa et al., 2024). Based on case studies, ionospheric SEDs are usually found to occur during the main phase of magnetic storms (DST<-50 nT, KP>5) and/or under prolonged southward IMF Bz and increased solar wind speed (e.g., Yizengaw et al., 2006; Liu et al., 2015; Zou et al., 2021; Li et al., 2022). Plasmaspheric plumes often occur during southward IMF Bz (Goldstein et al., 2003; 2004; Walsh et al., 2014), and during moderate storm, with 24-hour minimum values of Kp~3-6 or Dst between -70 and -10 nT (Darrouzet et al., 2008; Li et al., 2022). However, timing differences between these phenomena may be on the order of a day, making it difficult to determine the driver unambiguously.

### 1.3 Motivation and contribution

We identify two critical unsolved questions, which we address statistically in this study:

(1) Do the plasmaspheric plume and ionospheric SED plumes start and end concurrently, within measurable temporal uncertainty?

(2) What solar wind driving and geomagnetic activity are responsible for the onset of both plasmaspheric and SED plumes?

Establishing the relationship between the start and end times of plasmaspheric plumes and ionospheric SED plumes is critical for advancing our understanding of their generation mechanisms and magnetosphere-ionosphere coupling, and for improving operational space weather forecasting. Plasmaspheric plumes can mass-load the dayside magnetopause and reduce reconnection efficiency (Walsh et al., 2013, 2014, 2021), and they can also influence radiation belt dynamics through cold plasma influence on wave-particle interactions (Thorne et al., 2013; Ripoll et al., 2019; Ma et al., 2018). SEDs degrade GNSS performance by introducing steep TEC and electron density gradients and irregularities that distort GNSS radio-wave propagation through the ionosphere (Ledvina et al., 2002; Basu et al., 2005; Foster et al., 2005; Kintner et al., 2007; Sun et al., 2013). A confirmed strong one-to-one temporal correspondence between the start and end times of these structures over a large observation set provides key information to enable more accurate real-time monitoring and prediction of storm-time geospace disturbances.

In this study, we introduce a plasmapause asymmetry (PPA) index derived from the machine-learning-based DEN3D model of three-dimensional plasmaspheric electron density (Chu et al., 2017a, b). The PPA index (described below) enables systematic identification of plume intervals throughout geomagnetic storms. By comparing plume intervals with ionospheric SEDs identified from global TEC maps, we show that plasmaspheric plumes and SED plumes evolve in a one-to-one temporal correspondence, with their start and end times aligned to within roughly half an hour. A superposed epoch analysis further demonstrates that SED onsets coincide with peaks in IMF Bz and Ey, AE/AL indices, and ASY-H/D indices. These results establish that plume formation and SED enhancements are dynamically coupled on a consistent basis via magnetosphere-ionosphere electrodynamics, providing new physical insights and potential pathways for nowcasting and forecasting magnetospheric plasma dynamics and navigation disturbances.

## 2 Data and Methods

### 2.1 Data Description

This study utilizes three primary datasets:

(1) Ionospheric vertical TEC data at 1°×1° spatial resolution and 5-minute cadence, derived from large networks of ground-based GNSS receivers and obtained from the Madrigal database (Rideout & Coster, 2006; Vierinen et al., 2016);

(2) Solar wind and geomagnetic indices at 1-minute resolution obtained from the OMNI (Papitashvili et al., 2020) and SuperMAG (Gjerloev et al., 2012) databases;

(3) Plasmaspheric electron density distribution and evolution reconstructed using the neural-network-based DEN3D model (Chu et al., 2017a,b).

### 2.2 Plasmapause Asymmetry (PPA) Index

In order to facilitate our analysis, we define a plasmaspheric plasmapause asymmetry (PPA) index based on outputs from the DEN3D model as follows:

1. At each time $t_0$, the DEN3D model takes the time history of solar wind and geomagnetic indices as input to predict equatorial plasmaspheric density.

2. The nominal plasmapause location ($L_{pp}$) is extracted at all magnetic local times (MLT) as the plasma density contour at 50 $cm^{-3}$, which is the median density at the midpoint of the plasmapause (Denton et al., 2024). We also extract $L_{pp}$ using the 30 and 70 $cm^{-3}$ contours; their mean closely matches the 50 $cm^{-3}$ plasmapause location, providing both a consistency check and an uncertainty range.

3. The PPA index is calculated as the difference between the maximum and minimum $L_{pp}$ across MLTs. It represents the degree of plasmapause asymmetry, which serves as a proxy for plume formation, and ranges from ~0 (representing a symmetric plasmapause) to “high” values such as ~3-4 (representing significant plasmapause distortion).

4. The PPA index increases when the plasmapause extends outward in the noon-to-afternoon sector and erodes on the nightside, which is characteristic of plume formation (see Figure 6 in Chu et al., 2017a, Section 3, and the event analysis in Figure 1).

5. This calculation is performed at a 1-minute cadence, enabling high-resolution tracking of storm-time plasmasphere dynamics.

Using this method, we generated a 1-minute-resolution PPA index for magnetic storms between 2010-2024. Plasmaspheric plumes are identified when: (1) the PPA index exceeds 2.0 (peak), (2) start and end times are defined at a threshold of 1.8, (3) each event lasts at least 2 hours, and (4) no two plumes occur within a 6-hour window.

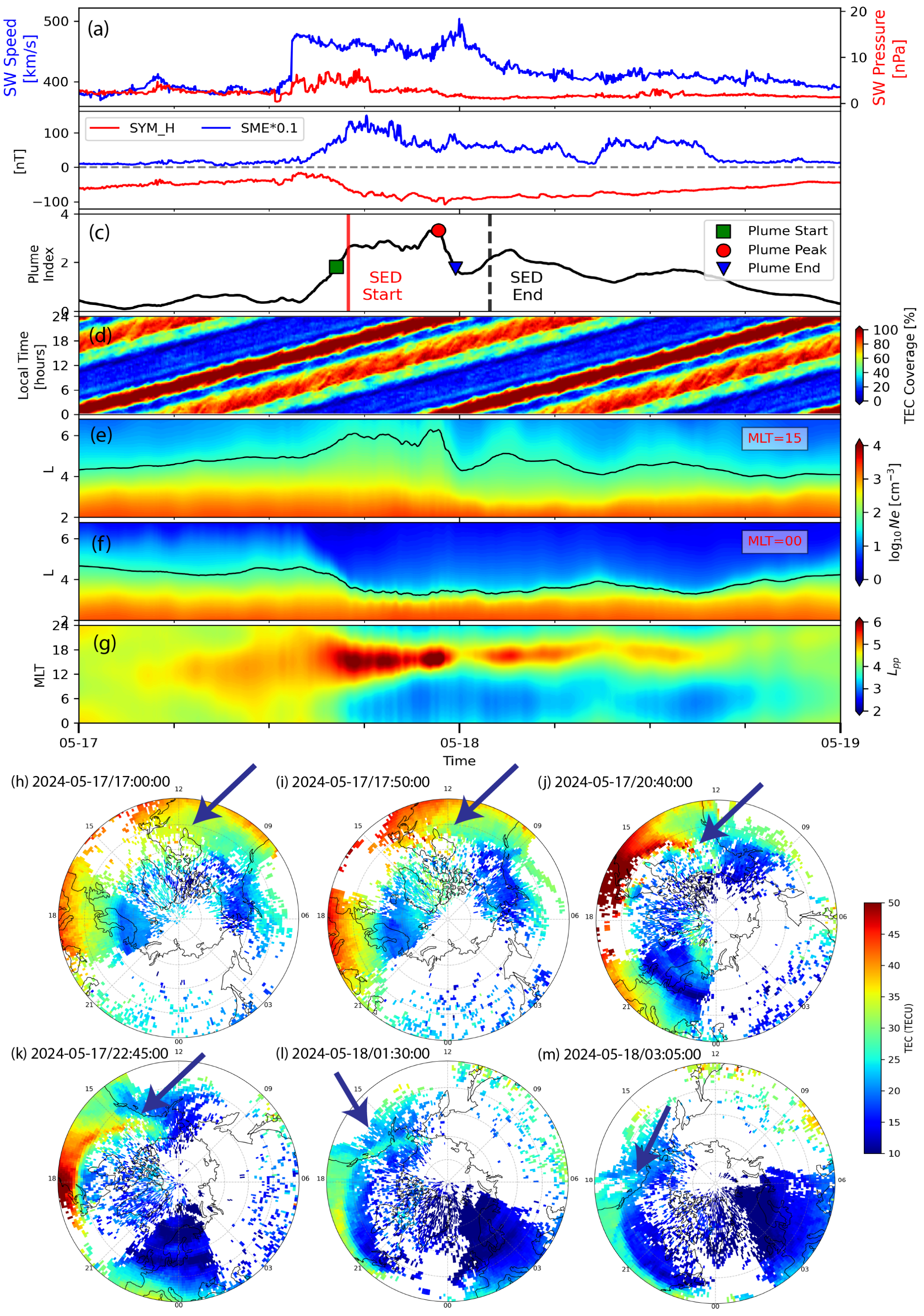

Figure 1. Event analysis demonstrating the SED plume and the deviation of the plasmaspheric plasmapause asymmetry (PPA) index during the magnetic storm on 17-19 May 2024. (a) Solar wind speed (blue) and pressure (red) from OMNI. (b) The SuperMAG equivalent AE index (blue) and the Sym-H index (red). (c) PPA index. (d) Percentage of TEC availability of 1°x1° grid between LAT=[40°, 60°] versus local time. (e-f) Predicted plasmaspheric density versus L shell at the afternoon (MLT=15) and midnight (MLT=0) meridian. Thin black lines represent the nominal plasmapause location (n=50 $cm^{-3}$). (g) Plasmapause location versus MLT. (h-m) Global TEC maps showing the appearance, sub-corotation, and disappearance of the storm-enhanced density. Black arrows in each panel highlight the SED plume location.

### 2.3 Ionospheric SED Events

We identify ionospheric SED plume events using the spatial and temporal evolution of global TEC maps during magnetic storms between 2010 and 2024, following the procedure described by Aa et al. (2024), when a distinct plume structure extends sunward/poleward with substantial TEC enhancements and gradients, where TEC values are doubled or even tripled compared to their poleward and equatorward counterparts (see Figure 1 in Aa et al., 2024). There are 92 geomagnetic storms between 2010 and 2024 with a minimum Sym-H index below -75 *nT*. This threshold was extended from -100 nT criterion used in Aa et al., 2024 in order to obtain a larger and more inclusive list for the present statistical analysis. Of these storms, we identify 75 storms in which clear SED plumes are visible. The remaining 17 storms lack sufficient TEC coverage in the noon-to-afternoon sector, where SEDs usually occurs; therefore, SEDs cannot be identified. The uncertainty in SED start and end times is approximately 20 min (estimated from 4 frames at TEC's 5 min cadence). Due to limited GNSS station coverage in some longitudinal sectors, some SED start or end times are based on an observation of the plume rotating into or out of the TEC field of view (FOV), rather than the actual times of appearance or disappearance of the SED.

## 3. Results and Discussion

### 3.1 Event Analysis of PPA Index

Figure 1 presents an event analysis of the geomagnetic storm on 17 May 2024, driven by a Coronal Mass Ejection (CME). Upon CME arrival at 13:25 UT, the solar wind speed increased rapidly from ~400 to ~500 km/s. The IMF Bz turned southward, reaching a minimum of -15 *nT*, and gradually returned toward zero over the next day. Sym-H reached a minimum of -110 *nT* near 23:00 UT, while the AE index peaked at ~1400 *nT* around 18:30 UT.

Figures 1e and 1f show the DEN3D-predicted plasmaspheric densities along the MLT=15 (post-noon) and MLT=0 (midnight) meridians. During the storm's main phase, the nightside plasmasphere was eroded, with plasma densities decreasing at L>3.0 and the 50 $cm^{-3}$ plasmapause contracting from $L_{pp}$=4.5 to $L_{pp}$=3.0. Conversely, in the post-noon sector, the plasmaspheric density increased, and the plasmapause extended outward from $L_{pp}$=5.0 to $L_{pp}$=6.0, indicating plume formation.

Figure 1g shows the full MLT distribution of $L_{pp}$, highlighting this asymmetric expansion. The corresponding PPA index (Figure 1c) began to rise as the nightside plasmapause contracted and the dayside plume emerged. A plasmaspheric plume is identified when the PPA index exceeds 2.0, with start and end times defined at the 1.8 threshold (marked by circles). In this event, the plasmaspheric plume started at 16:15 UT and ended at 23:45 UT, lasting approximately 7.5 hours.

### 3.2 Event Analysis of SED plume

Figures 1h-1m show the temporal evolution of global TEC maps during the 17 May 2024 magnetic storm. The SED plume first appeared around 17:00 UT, near local noon, at a geographic latitude of ~50°N, along the eastern coast of Canada. The associated TEC gradually increased in this region, with the SED plume becoming most pronounced at 20:40 UT (Figure 1j), centered near MLT=14 and spanning geographic latitudes from 45° to 60°N.

As the storm progressed, the SED plume sub-corotated westward at a speed slower than Earth's rotation, gradually weakening and disappearing by ~03:05 UT on 18 May over western Canada (Figure 1l). Both the SED's onset and disappearance occurred within the GNSS TEC's FOV (Figures 1h-1m) and were corroborated by the TEC coverage percentage in Figure 1d (1°×1° grid, 40°-60° latitude versus local time). Therefore, the SED start and end times can be identified directly, without ambiguity introduced by the plume rotating into or out of the TEC map's FOV.

Vertical lines in Figure 1c mark the SED's duration. Notably, the SED and plasmaspheric plume durations, derived independently from TEC maps and the PPA index, respectively, are in close agreement, with an offset of less than one hour. After the plasmaspheric plume ended, the SED continued sub-corotating for an additional ~2 hours before fading from the TEC maps.

The PPA index remained elevated after the first plume and exhibited a secondary enhancement, suggesting the possible development of a second plasmaspheric plume and associated SED plume. However, insufficient TEC coverage precludes confirmation. Multiple plume-SED pairs are observed in other storms (e.g., 10 October 2024) and will be explored in a follow-up study.

### 3.3 Temporal Correlations

In this section, we analyze the temporal relationships between the start and end times of plasmaspheric plumes and ionospheric SED plumes during geomagnetic storms.

Figure 2a shows the temporal alignment of plasmaspheric plumes and SED plume events across 75 storms, with all times referenced to the time of the Sym-H minimum. The PPA index during each storm is represented by vertical colored lines, spanning from the onset of the storm main phase through the recovery phase. Thin horizontal lines mark the start and end times of plasmaspheric plumes. All magnetic storms exhibit associated plasmaspheric plumes, which typically occur during the main phase and may extend into the early recovery phase. In some cases, plumes reappear during the recovery phase and are usually associated with additional magnetic activity, consistent with previous studies (Li et al., 2022). SED plumes also occur mainly during the storm main phase, with some occurring during the recovery phase. All 75 magnetic storms exhibit both plasmaspheric and SED plume signatures, demonstrating a strong temporal correlation between the two phenomena.

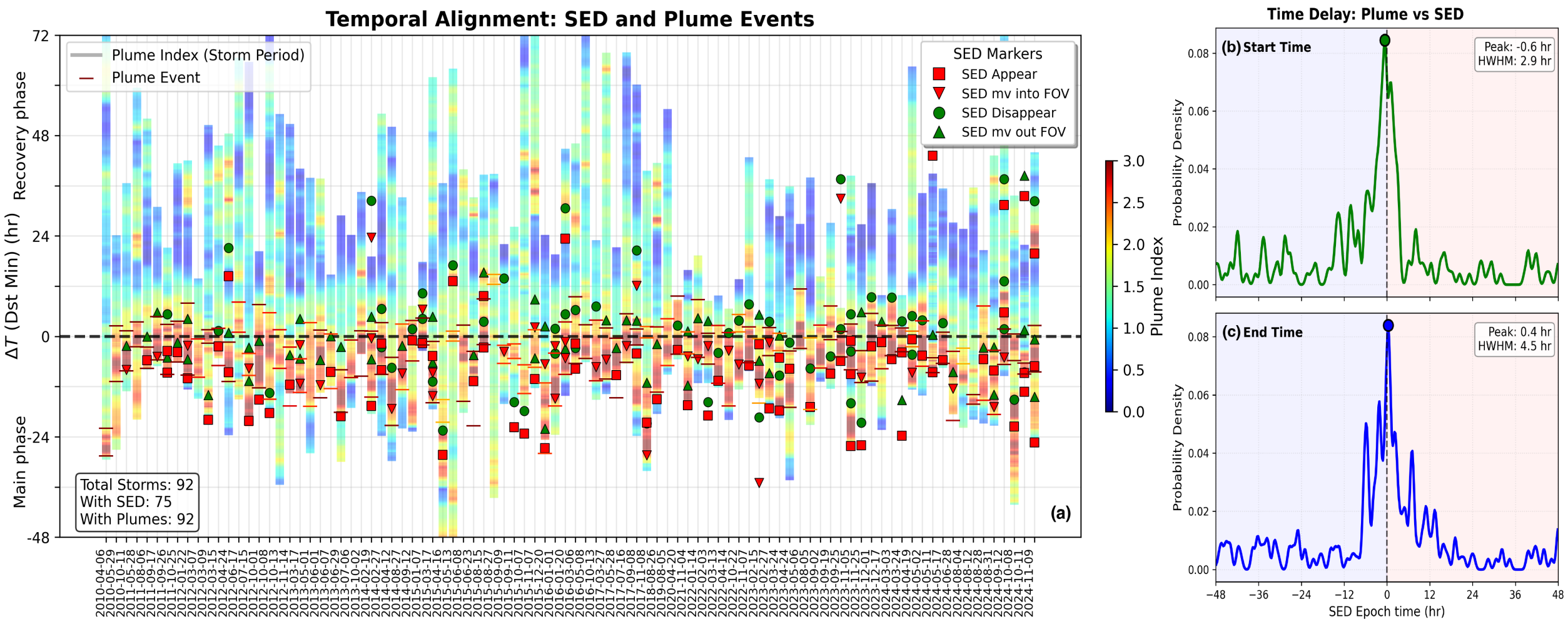

Figure 2. (a) Temporal alignment between the duration of the plasmaspheric plume and ionospheric SED plume relative to the minimum Sym-H of 92 magnetic storms between 2010 and 2024. The plume index is indicated by vertical colored lines, with epoch zero taken at the time of the minimum Sym-H for each storm. The start and end of each colored line mark the start of the main phase and the end of the recovery phase. The thin horizontal lines on each colored line indicate the duration of plasmaspheric plumes. The duration of SED plumes is marked by various symbols indicating their (dis)appearance and rotation into (out of) the FOV of TEC stations. (b-c) Time-delay distributions between the start and end times of the plasmaspheric plume and the ionospheric SEDs are shown in the right panels.

Figures 2b-2c show the time-delay distributions between the start and end times of plasmaspheric and SED plumes, with the SED times as the zero-epoch reference. Note that the times when SEDs rotate into or out of TEC's FOV are excluded to ensure accurate timing. Figure 2b shows a near-zero time delay (-0.6±2.9 hr) between their start times, suggesting that they occur nearly concurrently during magnetic storms. The time-delay distribution for their end times also centers around zero (0.4±4.5 hr) with a slightly broader spread. In addition, the temporal correlation appears stronger in more recent years (Figure 2a), likely due to improvements in TEC's observational coverage. These strong temporal correlations suggest that the plasmaspheric plumes and SED plumes usually appear and disappear concurrently within a few hours of each other during magnetic storms.

## 3.4 Solar Wind Driving and Geomagnetic Indices

In this section, we investigate the solar wind driving and geomagnetic conditions that trigger plasmaspheric plumes and ionospheric SED plumes. Using a comprehensive list of plume and SED start times, we perform superposed epoch analysis (SEA) of key solar wind and geomagnetic parameters relative to those onsets.

Figure 3 shows that plasmaspheric plume onsets occur under a clear and consistent pattern in solar wind and geomagnetic conditions. The IMF Bz, solar wind electric field Ey, and PC index all exhibit prominent peaks at plume onset, marking a transition from strong to weakening solar wind and convection electric fields. Solar-wind-magnetosphere coupling functions, including the universal coupling function (UCF; Newell et al., 2007) and the optimum coupling function (OPN; McPherron et al., 2015; Chu et al., 2024), also peak at plume onset, indicating that plume formation occurs as solar wind energy input and flux transfer into the magnetosphere reach a peak and then begin to weaken. Concurrently, SME reaches its maximum at plume onset, reflecting strong substorm-related magnetospheric convection and fast plasma flows that perturb the inner-magnetospheric electric field. The sharp decrease in Sym-H confirms that plume formation predominantly occurs during the storm main phase. The peak in the asymmetric ring current Asym-H index indicates an enhanced partial ring current, which is commonly associated with penetration electric fields and strong SAPS electric fields (Kunduri et al., 2018), which shapes the boundary of plasmaspheric plumes (Foster et al., 2002; Goldstein et al., 2003; 2005). In summary, these results show that plumes form during the transition from intense storm-time driving to a weakening convection state, under conditions of strong substorm- and SAPS-related electric fields.

SED onset exhibits similar temporal pattern across all solar wind and geomagnetic parameters. The SED onset aligns with pronounced peaks in IMF Bz, solar wind Ey, the PC index, and the solar-wind-magnetosphere coupling functions. These signatures indicate that SEDs form during the transition from enhanced to declining solar wind energy input and flux transfer and weakening convection. The sharp Sym-H drop confirms that SEDs initiate during the storm main phase, similar to plumes. SME and Asy-H peaks near the SED onset with small lags (1.1 hr for SME and 1.7 hr for Asy-H), consistent with their association with substorm-related convection, the partial ring current, and SAPS electric fields. Although time shifts appear in the SEA, they should not be interpreted as significant timing differences between SED and plasmaspheric plume onset due to the uncertainty along time axis (~hours). These differences may arise naturally from variations in event counts, solar-cycle distribution, event amplitudes, and statistical smoothing inherent in the SEA analysis.

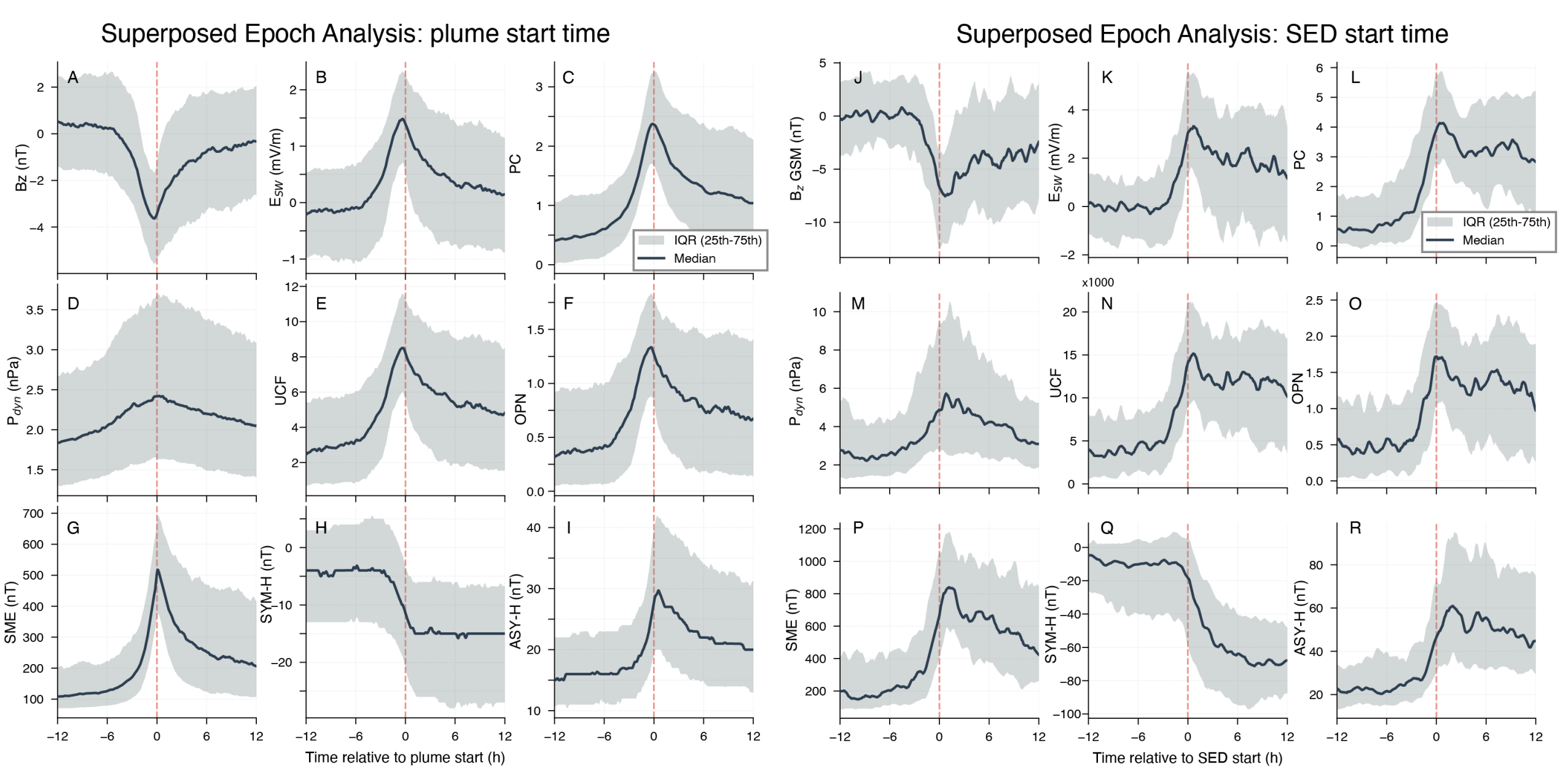

Figure 3. Superposed epoch analysis of solar wind parameters and geomagnetic indices relative to the start times of plasmaspheric plumes and ionospheric SED plumes.

**4. Conclusion**

This study demonstrates the existence of a strong temporal correlation between plasmaspheric plumes and ionospheric SED plumes during geomagnetic storms. Using the DEN3D neural-network-based electron density model, we developed a plasmapause asymmetry (PPA) index and identified plasmaspheric plumes in the period ranging from 2010 to 2024. In parallel, we identify SED intervals from global TEC maps for 92 magnetic storms. Of these storms, 75 had continuous and sufficient TEC coverage in the afternoon sector to enable reliable SED detection. All 75 storms exhibited coincident SED and plasmaspheric plumes, establishing a robust one-to-one correspondence (100%) except for these storms without observable SEDs due to observed TEC gaps.

A key result is the strong temporal alignment that was found between the start and end times of plasmaspheric plumes and ionospheric SED plumes. The time-lag distributions for both the start and end times are sharply peaked at zero epoch, indicating that the two phenomena typically appear and disappear nearly concurrently. Although SED and plasmaspheric plumes are generated by different physical processes and in different regions, their synchronized timing suggests that they respond to the same drivers and highlights the essential role of magnetosphere-ionosphere coupling under storm-time conditions. This coupling is most likely mediated by the rapid reconfiguration of electric fields and potentials mapped along magnetic field lines (Foster et al., 2002; Goldstein et al., 2003; 2005). These results provide the first long-term (2010-2024) statistical validation of the temporal correlation between the start and end times of these two key storm-time phenomena in the geospace plume after the IMAGE era, aided by a machine-learning-based DEN3D model. Their spatial relationship will be investigated in future work.

The superposed epoch analysis reveals the solar wind and magnetospheric drivers that are potentially responsible for generating SEDs and plasmaspheric plumes. Plasmaspheric plume onsets occur during a coherent sequence of magnetospheric driving: enhanced storm-time convection strengthens the dawn-dusk electric field, causing the Alfvén layer to contract and erode the nightside plasmasphere. Cold plasma drifts along reconfigured equipotentials toward the dayside magnetopause, forming a sunward-directed plasmaspheric plume. This process is supported by three primary electrodynamic contributions: (1) the direct-driven electric field, indicated by the peak in the PC index; (2) the substorm-related convection electric field from nightside reconnection, indicated by the peak in the SME index; and (3) the SAPS electric field, which sharpens the eastward edge of plasmaspheric plumes (Goldstein et al., 2003). The SAPS electric field arises from an enhanced partial ring current in the afternoon sector, indicated by the peak in the Asym-H index. In contrast, ionospheric SEDs are believed to be generated by (1) the storm-time penetration electric field, which increases mid-latitude plasma densities and redirects plasma poleward (Heelis et al., 2009; Manoj and Maus, 2012), supported by peaks in solar wind Ey, PC, and coupling functions, and (2) the SAPS electric field within the Region-2 current system (Foster et al., 2007), supported by peaks in the partial ring current typically associated with SAPS. In summary, the SEA results support generation mechanisms for both plasmaspheric and SED plumes within a common electrodynamic framework established by storm-time variations in electric fields and potentials mapped along magnetosphere-ionosphere field lines: (1) convection and penetration electric fields driven by the solar wind, (2) substorm-related convection electric fields on the nightside, and (3) SAPS electric fields associated with the partial ring current.

The PPA index provides a powerful diagnostic for space weather research and operational forecasting. Plasmaspheric density and plume structure strongly influence radiation belt dynamics, particularly in regions where wave-particle interactions can accelerate or scatter relativistic electrons. Cold plasma also modulates dayside magnetic reconnection efficiency through mass loading by heavy ions. With the loss of global plasmaspheric imaging following the decommissioning of the IMAGE mission, the DEN3D model and the PPA index provide an effective surrogate for continuous monitoring and forecasting of plasmaspheric dynamics. In future work, the PPA index could be extended from a diagnostic based on DEN3D to a nowcasting and forecasting capability by deriving it from operational numerical models, such as WAM-IPE, that provide time-dependent plasmaspheric density distributions.

Moreover, the PPA index may help identify SEDs during magnetic storms, thereby supporting ionospheric research and operational space weather applications, particularly since the IMAGE satellite was decommissioned. Because SEDs degrade GNSS performance through sharp TEC gradients and associated small-scale irregularities, the PPA index serves as a valuable early indicator of conditions that may lead to navigation and communication disruptions. More broadly, this work highlights the transformative potential of machine learning in space physics by enabling new diagnostics of coupled geospace phenomena that were previously inaccessible due to observational limitations.

**Acknowledgments**

XC, JG, JB, NM, KC would like to thank grant 80NSSC22K1023, 80NSSC23K0096, 80NSSC24K1112, NSF grant AGS-2247255, and AFOSR YIP FA9550-23-1-0359. In addition, JB acknowledges support from subgrants to the University of California, Los Angeles, from the University of Colorado Boulder under NASA Prime Grant agreements 80NSSC20K1580 and 80NSSC22K1023, and from NSF award AGS-2025706. This work uses the Casper system provided by the NSF National Center for Atmospheric Research (NCAR), sponsored by the National Science Foundation and Extreme Science and Engineering Discovery Environment (XSEDE) Bridges GPU at the PSC through allocation TG-PHY190033. GNSS TEC data products and access through the Madrigal distributed data system are provided to the community by the Massachusetts Institute of Technology under support from US National Science Foundation grants AGS-1242204, AGS-1952737, and AGS-2504079. We gratefully acknowledge useful discussions with Ercha Aa at MIT Haystack Observatory, and with the NASA LWS team.

**Open Research**

The GNSS TEC data product is obtained through the Madrigal community database operated by the Millstone Hill Geospace Facility (https://cedar.openmadrigal.org/). Solar wind parameters and geomagnetic indices are sourced from the OMNI dataset (https://omniweb.gsfc.nasa.gov) and from SuperMAG (https://supermag.jhuapl.edu). The neural network models were developed using the open-source TensorFlow package (https://www.tensorflow.org).

## Conflict of Interest Disclosure

The authors declare there are no conflicts of interest for this manuscript.

## References

Aa, E., Dzwill, P., Zhang, S.-R., & Erickson, P. J. (2024). A Statistical Analysis of the Morphology of Storm-Enhanced Density Plumes Over the North American Sector. *Journal of Geophysical Research: Space Physics*, *129*(6), e2024JA032750. https://doi.org/10.1029/2024JA032750

Aa, E., Zhang, S.-R., Wang, W., Erickson, P. J., & Coster, A. J. (2023). Multiple Longitude Sector Storm-Enhanced Density (SED) and Long-Lasting Subauroral Polarization Stream (SAPS) During the 26–28 February 2023 Geomagnetic Storm. *Journal of Geophysical Research: Space Physics*, *128*(9), e2023JA031815. https://doi.org/10.1029/2023JA031815

Anderson, D. N. (1976). Modeling the midlatitude *F*-region ionospheric storm using east-west drift and a meridional wind. *Planetary and Space Science*, *24*(1), 69–77. https://doi.org/10.1016/0032-0633(76)90063-5

Balan, N., Shiokawa, K., Otsuka, Y., Kikuchi, T., Vijaya Lekshmi, D., Kawamura, S., Yamamoto, M., & Bailey, G. J. (2010). A physical mechanism of positive ionospheric storms at low latitudes and midlatitudes. *Journal of Geophysical Research: Space Physics*, *115*(A2). https://doi.org/10.1029/2009JA014515

Basu, Su., Basu, S., Makela, J. J., Sheehan, R. E., MacKenzie, E., Doherty, P., Wright, J. W., Keskinen, M. J., Pallamraju, D., Paxton, L. J., & Berkey, F. T. (2005). Two components of ionospheric plasma structuring at midlatitudes observed during the large magnetic storm of October 30, 2003. *Geophysical Research Letters*, *32*(12). https://doi.org/10.1029/2004GL021669

Chen, A. J., & Grebowsky, J. M. (1974). Plasma tail interpretations of pronounced detached plasma regions measured by Ogo 5. *Journal of Geophysical Research (1896-1977)*, *79*(25), 3851–3855. https://doi.org/10.1029/JA079i025p03851

Chu, X., Bortnik, J., Li, W., Ma, Q., Denton, R., Yue, C., Angelopoulos, V., Thorne, R. M., Darrouzet, F., Ozhogin, P., Kletzing, C. A., Wang, Y., & Menietti, J. (2017). A neural network model of three-dimensional dynamic electron density in the inner magnetosphere. *Journal of*

*Geophysical Research: Space Physics*, *122*(9), 9183–9197. https://doi.org/10.1002/2017JA024464

Chu, X. N., Bortnik, J., Li, W., Ma, Q., Angelopoulos, V., & Thorne, R. M. (2017). Erosion and refilling of the plasmasphere during a geomagnetic storm modeled by a neural network. *Journal of Geophysical Research: Space Physics*, *122*(7), 7118–7129. https://doi.org/10.1002/2017JA023948

Coster, A. J., Erickson, P. J., Foster, J. C., Thomas, E. G., Ruohoniemi, J. M., & Baker, J. (2016). Solar Cycle 24 Observations of Storm-Enhanced Density and the Tongue of Ionization. In *Ionospheric Space Weather* (pp. 71–83). American Geophysical Union (AGU). https://doi.org/10.1002/9781118929216.ch6

Darrouzet, F., De Keyser, J., Décréau, P. M. E., El Lemdani-Mazouz, F., & Vallières, X. (2008). Statistical analysis of plasmaspheric plumes with Cluster/WHISPER observations. *Annales Geophysicae*, *26*(8), 2403–2417. https://doi.org/10.5194/angeo-26-2403-2008

Foster, J. C. (1993). Storm time plasma transport at middle and high latitudes. *Journal of Geophysical Research: Space Physics*, *98*(A2), 1675–1689. https://doi.org/10.1029/92JA02032

Foster, J. C., Coster, A. J., Erickson, P. J., Holt, J. M., Lind, F. D., Rideout, W., McCready, M., van Eyken, A., Barnes, R. J., Greenwald, R. A., & Rich, F. J. (2005). Multiradar observations of the polar tongue of ionization. *Journal of Geophysical Research: Space Physics*, *110*(A9). https://doi.org/10.1029/2004JA010928

Foster, J. C., Erickson, P. J., Coster, A. J., Goldstein, J., & Rich, F. J. (2002). Ionospheric signatures of plasmaspheric tails. *Geophysical Research Letters*, *29*(13), 1-1-1–4. https://doi.org/10.1029/2002GL015067

Foster, J. C., Erickson, P. J., Walsh, B. M., Wygant, J. R., Coster, A. J., & Zhang, Q.-H. (2020). Multi-Point Observations of the Geospace Plume. In *Dayside Magnetosphere Interactions* (pp. 243–264). American Geophysical Union (AGU). https://doi.org/10.1002/9781119509592.ch14

Foster, J. C., & Rideout, W. (2005). Midlatitude TEC enhancements during the October 2003 superstorm. *Geophysical Research Letters*, *32*(12). https://doi.org/10.1029/2004GL021719

Foster, J. C., Rideout, W., Sandel, B., Forrester, W. T., & Rich, F. J. (2007). On the relationship of SAPS to storm-enhanced density. *Journal of Atmospheric and Solar-Terrestrial Physics*, *69*(3), 303–313. https://doi.org/10.1016/j.jastp.2006.07.021

Foster, J. C., & Erickson, P. J. (2013). Ionospheric superstorms: Polarization terminator effects in the Atlantic sector. Journal of Atmospheric and Solar-Terrestrial Physics, 103, 147-156. https://doi.org/10.1016/j.jastp.2013.04.001

Foster, J.C., Zou, S., Heelis, R.A. and Erickson, P.J. (2021). Ionospheric Storm-Enhanced Density Plumes. In Ionosphere Dynamics and Applications (eds C. Huang, G. Lu, Y. Zhang and L.J.

Paxton). https://doi.org/10.1002/9781119815617.ch6

Garcia, L. N., Fung, S. F., Green, J. L., Boardsen, S. A., Sandel, B. R., & Reinisch, B. W. (2003). Observations of the latitudinal structure of plasmaspheric convection plumes by IMAGE-RPI and EUV. *Journal of Geophysical Research: Space Physics*, *108*(A8). https://doi.org/10.1029/2002JA009496

Gjerloev, J. W. (2012). The SuperMAG data processing technique. *Journal of Geophysical Research: Space Physics*, *117*(A9). https://doi.org/10.1029/2012JA017683

Goldstein, J., Burch, J. L., & Sandel, B. R. (2005). Magnetospheric model of subauroral polarization stream. *Journal of Geophysical Research: Space Physics*, *110*(A9). https://doi.org/10.1029/2005JA011135

Goldstein, J., Sandel, B. R., Hairston, M. R., & Reiff, P. H. (2003). Control of plasmaspheric dynamics by both convection and sub-auroral polarization stream. *Geophysical Research Letters*, *30*(24). https://doi.org/10.1029/2003GL018390

Goldstein, J., Spasojević, M., Reiff, P. H., Sandel, B. R., Forrester, W. T., Gallagher, D. L., & Reinisch, B. W. (2003). Identifying the plasmapause in IMAGE EUV data using IMAGE RPI in situ steep density gradients. *Journal of Geophysical Research-Space Physics*, *108*(A4), 1147. https://doi.org/10.1029/2002ja009475

Heelis, R. A., Sojka, J. J., David, M., & Schunk, R. W. (2009). Storm time density enhancements in the middle-latitude dayside ionosphere. *Journal of Geophysical Research: Space Physics*, *114*(A3). https://doi.org/10.1029/2008JA013690

Immel, T. J., Crowley, G., Craven, J. D., & Roble, R. G. (2001). Dayside enhancements of thermospheric O/N2 following magnetic storm onset. *Journal of Geophysical Research: Space Physics*, *106*(A8), 15471–15488. https://doi.org/10.1029/2000JA000096

J. Goldstein, Jerry Goldstein, Goldstein, J., Goldstein, J., J. L. Burch, Burch, J. L., B. R. Sandel, & Sandel, B. R. (2005). Magnetospheric model of subauroral polarization stream. *Journal of Geophysical Research*, *110*. https://doi.org/10.1029/2005ja011135

Kelley, M. C., Vlasov, M. N., Foster, J. C., & Coster, A. J. (2004). A quantitative explanation for the phenomenon known as storm-enhanced density. *Geophysical Research Letters*, *31*(19). https://doi.org/10.1029/2004GL020875

Ledvina, B. M., Makela, J. J., & Kintner, P. M. (2002). First observations of intense GPS L1 amplitude scintillations at midlatitude. *Geophysical Research Letters*, *29*(14), 4-1-4–4. https://doi.org/10.1029/2002GL014770

Li, B., Le, H., Li, W., Chen, Y., & Liu, L. (2022). Longitudinal Evolution of Storm-Enhanced

Densities: A Case Study. *Remote Sensing*, *14*(24), 6340. https://doi.org/10.3390/rs14246340

Li, H., Fu, T., Tang, R., Yuan, Z., Yang, Z., Ouyang, Z., & Deng, X. (2022). Statistical study and corresponding evolution of plasmaspheric plumes under different levels of geomagnetic storms. *Annales Geophysicae*, *40*(2), 167–177. https://doi.org/10.5194/angeo-40-167-2022

Li, S., Liu, J., Wang, W., Liang, J., & Zhang, K. (2023). Impacts of Subauroral Polarization Streams on Storm-Enhanced Density Plume and Consequently on Polar Tongue of Ionization. *Earth and Space Science*, *10*(9), e2023EA002827. https://doi.org/10.1029/2023EA002827

Liu, J., Wang, W., Burns, A., Solomon, S. C., Zhang, S., Zhang, Y., & Huang, C. (2016). Relative importance of horizontal and vertical transports to the formation of ionospheric storm-enhanced density and polar tongue of ionization. *Journal of Geophysical Research: Space Physics*, *121*(8), 8121–8133. https://doi.org/10.1002/2016JA022882

Liu, J., Wang, W., Burns, A., Yue, X., Zhang, S., Zhang, Y., & Huang, C. (2016). Profiles of ionospheric storm-enhanced density during the 17 March 2015 great storm. *Journal of Geophysical Research: Space Physics*, *121*(1), 727–744. https://doi.org/10.1002/2015JA021832

Ma, Q., Li, W., Bortnik, J., Thorne, R. M., Chu, X., Ozeke, L. G., Reeves, G. D., Kletzing, C. A., Kurth, W. S., Hospodarsky, G. B., Engebretson, M. J., Spence, H. E., Baker, D. N., Blake, J. B., Fennell, J. F., & Claudepierre, S. G. (2018). Quantitative Evaluation of Radial Diffusion and Local Acceleration Processes During GEM Challenge Events. *Journal of Geophysical Research: Space Physics*, *123*(3), 1938–1952. https://doi.org/10.1002/2017JA025114

Mannucci, A. J., Tsurutani, B. T., Iijima, B. A., Komjathy, A., Saito, A., Gonzalez, W. D., Guarnieri, F. L., Kozyra, J. U., & Skoug, R. (2005). Dayside global ionospheric response to the major interplanetary events of October 29–30, 2003 "Halloween Storms." *Geophysical Research Letters*, *32*(12). https://doi.org/10.1029/2004GL021467

Manoj, C., & Maus, S. (2012). A real-time forecast service for the ionospheric equatorial zonal electric field. *Space Weather*, *10*(9). https://doi.org/10.1029/2012SW000825

Mendillo, M. (2006). Storms in the ionosphere: Patterns and processes for total electron content. *Reviews of Geophysics*, *44*(4). https://doi.org/10.1029/2005RG000193

Moldwin, M. B., Howard, J., Sanny, J., Bocchicchio, J. D., Rassoul, H. K., & Anderson, R. R. (2004). Plasmaspheric plumes: CRRES observations of enhanced density beyond the plasmapause. *Journal of Geophysical Research: Space Physics*, *109*(A5). https://doi.org/10.1029/2003JA010320

Moldwin, M. B., Sandel, B. R., Thomsen, M. F., & Elphic, R. C. (2003). Quantifying Global Plasmaspheric Images With in situ Observations. *Space Science Reviews*, *109*(1), 47–61. https://doi.org/10.1023/B:SPAC.0000007512.69979.8f

Moldwin, M. B., Zou, S., & Heine, T. (2016). The story of plumes: The development of a new conceptual framework for understanding magnetosphere and ionosphere coupling. *Annales Geophysicae*, *34*(12), 1243–1253. https://doi.org/10.5194/angeo-34-1243-2016

Ober, D. M., Horwitz, J. L., Thomsen, M. F., Elphic, R. C., McComas, D. J., Belian, R. D., & Moldwin, M. B. (1997). Premidnight plasmaspheric "plumes." *Journal of Geophysical Research: Space Physics*, *102*(A6), 11325–11334. https://doi.org/10.1029/97JA00562

Papitashvili, N. E., & King, J. H. (2020). *OMNI 1-min data* [Dataset]. https://omniweb.gsfc.nasa.gov/

Rideout, W., & Coster, A. (2006). Automated GPS processing for global total electron content data. *GPS Solutions*, *10*(3), 219–228. https://doi.org/10.1007/s10291-006-0029-5

Ripoll, J.-F., Claudepierre, S. G., Ukhorskiy, A. Y., Colpitts, C., Li, X., Fennell, J. F., & Crabtree, C. (2020). Particle Dynamics in the Earth's Radiation Belts: Review of Current Research and Open Questions. *Journal of Geophysical Research: Space Physics*, *125*(5), e2019JA026735. https://doi.org/10.1029/2019JA026735

Sandel, B. R., King, R. A., Forrester, W. T., Gallagher, D. L., Broadfoot, A. L., & Curtis, C. C. (2001). Initial results from the IMAGE Extreme Ultraviolet Imager. *Geophysical Research Letters*, *28*(8), 1439–1442. https://doi.org/10.1029/2001GL012885

Su, Y.-J., Thomsen, M. F., Borovsky, J. E., & Foster, J. C. (2001). A linkage between polar patches and plasmaspheric drainage plumes. *Geophysical Research Letters*, *28*(1), 111–113. https://doi.org/10.1029/2000GL012042

Sun, Y.-Y., Matsuo, T., Araujo-Pradere, E. A., & Liu, J.-Y. (2013). Ground-based GPS observation of SED-associated irregularities over CONUS. *Journal of Geophysical Research: Space Physics*, *118*(5), 2478–2489. https://doi.org/10.1029/2012JA018103

Thorne, R. M., Li, W., Ni, B., Ma, Q., Bortnik, J., Chen, L., Baker, D. N., Spence, H. E., Reeves, G. D., Henderson, M. G., Kletzing, C. A., Kurth, W. S., Hospodarsky, G. B., Blake, J. B., Fennell, J. F., Claudepierre, S. G., & Kanekal, S. G. (2013). Rapid local acceleration of relativistic radiation-belt electrons by magnetospheric chorus. *Nature*, *504*(7480), 411–414. https://doi.org/10.1038/nature12889

Tsurutani, B., Mannucci, A., Iijima, B., Abdu, M. A., Sobral, J. H. A., Gonzalez, W., Guarnieri, F., Tsuda, T., Saito, A., Yumoto, K., Fejer, B., Fuller-Rowell, T. J., Kozyra, J., Foster, J. C., Coster, A., & Vasyliunas, V. M. (2004). Global dayside ionospheric uplift and enhancement associated with interplanetary electric fields. *Journal of Geophysical Research: Space Physics*, *109*(A8). https://doi.org/10.1029/2003JA010342

Vierinen, J., Coster, A. J., Rideout, W. C., Erickson, P. J., & Norberg, J. (2016). Statistical framework for estimating GNSS bias. *Atmospheric Measurement Techniques*, *9*(3), 1303–1312.

https://doi.org/10.5194/amt-9-1303-2016

Walsh, B. M., Phan, T. D., Sibeck, D. G., & Souza, V. M. (2014). The plasmaspheric plume and magnetopause reconnection. *Geophysical Research Letters*, *41*(2), 223–228. https://doi.org/10.1002/2013GL058802

Walsh, B. M., Sibeck, D. G., Nishimura, Y., & Angelopoulos, V. (2013). Statistical analysis of the plasmaspheric plume at the magnetopause. *Journal of Geophysical Research: Space Physics*, *118*(8), 4844–4851. https://doi.org/10.1002/jgra.50458

Walsh, B. M., & Zou, Y. (2021). The role of magnetospheric plasma in solar wind-magnetosphere coupling: A review. *Journal of Atmospheric and Solar-Terrestrial Physics*, *219*, 105644. https://doi.org/10.1016/j.jastp.2021.105644

Wang, W., Talaat, E. R., Burns, A. G., Emery, B., Hsieh, S., Lei, J., & Xu, J. (2012). Thermosphere and ionosphere response to subauroral polarization streams (SAPS): Model simulations. *Journal of Geophysical Research: Space Physics*, *117*(A7). https://doi.org/10.1029/2012JA017656

Yizengaw, E., Dewar, J., MacNeil, J., Moldwin, M. B., Galvan, D., Sanny, J., Berube, D., & Sandel, B. (2008). The occurrence of ionospheric signatures of plasmaspheric plumes over different longitudinal sectors. *Journal of Geophysical Research: Space Physics*, *113*(A8). https://doi.org/10.1029/2007JA012925

Yizengaw, E., Moldwin, M. B., & Galvan, D. A. (2006). Ionospheric signatures of a plasmaspheric plume over Europe. *Geophysical Research Letters*, *33*(17). https://doi.org/10.1029/2006GL026597

Zou, S., Moldwin, M. B., Ridley, A. J., Nicolls, M. J., Coster, A. J., Thomas, E. G., & Ruohoniemi, J. M. (2014). On the generation/decay of the storm-enhanced density plumes: Role of the convection flow and field-aligned ion flow. *Journal of Geophysical Research: Space Physics*, *119*(10), 8543–8559. https://doi.org/10.1002/2014JA020408

Zou, S., Perry, G. W., & Foster, J. C. (2021). Recent Advances in Polar Cap Density Structure Research. In *Ionosphere Dynamics and Applications* (pp. 67–82). American Geophysical Union (AGU). https://doi.org/10.1002/9781119815617.ch4

Zou, S., Ren, J., Wang, Z., Sun, H., & Chen, Y. (2021). Impact of Storm-Enhanced Density (SED) on Ion Upflow Fluxes During Geomagnetic Storm. *Frontiers in Astronomy and Space Sciences*, *8*. https://doi.org/10.3389/fspas.2021.746429